\documentclass[runningheads]{llncs}
\usepackage[utf8]{inputenc}
\usepackage[T1]{fontenc}
\usepackage[scaled=0.8]{beramono}

\newif\ifExtended
\Extendedtrue

\usepackage{wrapfig}
\usepackage{hyperref}
\usepackage{amsmath}

\usepackage[capitalize]{cleveref}
\usepackage{amssymb}
\usepackage{pifont}
\newcommand{\cmark}{\ding{51}}%
\newcommand{\xmark}{\ding{55}}%
\usepackage{stmaryrd}
\usepackage{color}
\usepackage{mathpartir}
\usepackage{array}
\usepackage{booktabs}
\usepackage{multicol}
\usepackage{todonotes}
\usepackage{comment}

\hypersetup{allcolors=blue,colorlinks=true}

\usepackage{listings}

\lstdefinestyle{customc}{
  escapechar=\%,
  frame=none,
  language=C,
  showstringspaces=false,
  basicstyle=\footnotesize\ttfamily,
  keywordstyle=\bfseries\color{blue},
  commentstyle=\slshape\color{red!50!black},
  identifierstyle=\color{black},
  stringstyle=\color{orange},
  morekeywords={assert,presumes,exists,fails,insecure,assertion,covers,size_t}
}

\usepackage{xspace}

\usepackage{tikz}

\definecolor{lcol}{rgb}{0,0,0.5}
\definecolor{errcolor}{rgb}{0.8,0,0}
\definecolor{inseccolor}{rgb}{0.7,0.4,0}
\definecolor{okcolor}{rgb}{0,0.7,0} 

\title{Compositional Vulnerability Detection with Insecurity Separation Logic\ifExtended\ (Extended Version)\fi}
\author{Toby Murray\inst{1}, Pengbo Yan\inst{1} and Gidon Ernst\inst{2}}
\institute{University of Melbourne, Melbourne, Australia \\
\email{toby.murray@unimelb.edu.au} \quad
\email{pengpoy@student.unimelb.edu.au}
\and
LMU Munich, Munich, Germany \\ \email{gidon.ernst@lmu.de}}

\newcommand{\InsecSL}{\textsc{InsecSL}\xspace}

\newcommand{\Tool}{\textsc{Underflow}\xspace}

\newcommand{\InferTool}{\textsc{Pulse-InsecSL}\xspace}

\let\star\ast

\newcommand{\SecCSL}{\textsc{SecCSL}\xspace}

\newcommand{\eval}[2]{\lbrack#1\rbrack_{#2}}

    \newcommand{\iprove}[4]{\vdash_{#1} \textcolor{blue}{[#2]}\ \textcolor{black}{#3}\ \textcolor{blue}{[\epsilon\!:\ #4]}}

  \newcommand{\iproveOK}[4]{\vdash_{#1} \textcolor{blue}{[#2]}\ \textcolor{black}{#3}\ \textcolor{okcolor}{[{\it ok\!:}\ #4]}}

  \newcommand{\iproveINSEC}[5]{\vdash_{#1} \textcolor{blue}{[#2]}\ \textcolor{black}{#3}\ \textcolor{inseccolor}{[{\it insec(#4)\!:}\ #5]}}

  \newcommand{\iproveERR}[5]{\vdash_{#1} \textcolor{blue}{[#2]}\ \textcolor{black}{#3}\ \textcolor{errcolor}{[{\it err(#4)\!:}\ #5]}}

  \newcommand{\low}{\kw{low}}
\newcommand{\high}{\kw{high}}  
\newcommand{\kw}[1]{\text{\bfseries\upshape #1}}
\newcommand{\While}[2]{\kw{while}\ #1\ \kw{do}\ #2\ \kw{done}}
\newcommand{\ITE}[3]{\kw{if}\ #1\ \kw{then}\ #2\ \kw{else}\ #3\ \kw{endif}}
\newcommand{\Assume}[1]{\kw{assume}(#1)}
\newcommand{\Skip}{\kw{skip}}
\newcommand{\Seq}[2]{#1; #2}
  \newcommand{\Assign}[2]{#1 := #2}
  \newcommand{\Load}[2]{\Assign{#1}{[#2]}}
  \newcommand{\Store}[2]{\Assign{[#1]}{#2}}
  \newcommand{\Input}[2]{\Assign{#1}{\kw{input}(#2)}}    

  \newcommand{\fv}[1]{\mathit{fv}(#1)}
  \newcommand{\Emp}{\textbf{emp}}
  \newcommand{\pto}[2]{#1 \mapsto #2}
  \newcommand{\pinvalid}[1]{#1 \not\mapsto}
  \newcommand{\union}{\cup}
  \newcommand{\inter}{\cap}
  \newcommand{\Label}[1]{\mathit{#1}\!:}
  \newcommand{\Alloc}[2]{\Assign{#1}{\kw{alloc}(#2)}}
  \newcommand{\Free}[1]{\kw{free}(#1)}
  \newcommand{\Output}[2]{\kw{output}(#1,#2)}
  \newcommand{\haslabel}{\mathit{::}}
  \newcommand{\Sec}[2]{#1 \mathbin{\haslabel} #2}
  \newcommand{\Insec}[2]{#1 \mathbin{\rlap{$\haslabel$}\hspace*{-0.9pt}{\not{~}}} #2}
    
  \newcommand{\mods}[1]{\mathit{mod}(#1)}
  \newcommand{\follows}[3]{#1 \stackrel{#2}{\Longrightarrow} #3}
  
  \newcommand{\true}{\kw{true}}
  
  \newcommand{\Val}{\mathit{Val}}
  \newcommand{\dom}[1]{\mathit{dom}(#1)}

  \newcommand{\Run}[3]{\langle\kw{run}\ \text{``}#1\text{''}\ #2\ #3\rangle}
  \newcommand{\Stop}[2]{\langle\kw{stop}\ #1\ #2\rangle}
  \newcommand{\Abort}[2]{\langle\kw{abort}\ #1\ #2\rangle}    

  \newcommand{\sems}[3]{#1 \xrightarrow{~#2~} #3}
  \newcommand{\semsstar}[3]{#1 {\xrightarrow{~#2~}}{\!}^* #3}

  \newcommand{\singlist}[1]{[#1]}
  
  \newcommand{\funupd}[3]{#1(#2:=#3)}
  \newcommand{\Out}[2]{\mathsf{out}\langle#1,#2\rangle}
  \newcommand{\Allocate}[1]{\mathsf{allocate}\langle#1\rangle}
  \newcommand{\In}[2]{\mathsf{in}\langle#1,#2\rangle}    

  \renewcommand{\subsubsection}[1]{\smallskip\noindent\textbf{#1}}
  
\begin{document}

\maketitle

\begin{abstract}
  Memory-safety issues and information leakage are known to be depressingly common. We consider the compositional static detection of these kinds of vulnerabilities in first-order C-like programs. Indeed the latter are \emph{relational}
  hyper-safety violations, comparing pairs of program executions, making them more challenging to detect than the former, which require reasoning only over individual executions. 
  Existing symbolic leakage detection methods
  treat only non-interactive programs, avoiding the challenges of nondeterminism. 
  Also, being whole-program analyses they cannot be applied
  one-function-at-a-time, thereby ruling out incremental analysis.
  We remedy these shortcomings by presenting Insecurity Separation Logic (\InsecSL), an under-approximate relational program logic for soundly detecting information leakage and memory-safety issues in interactive programs.
  Importantly, \InsecSL reasons about pairs of executions, and so is relational, but purposefully resembles the non-relational Incorrectness Separation Logic (ISL) that is already automated in the Infer tool.
  We show how \InsecSL can be automated by bi-abduction based symbolic execution,
  and we evaluate two implementations of this idea (one based on Infer)
  on various case-studies.
\end{abstract}

\section{Introduction}\label{sec:intro}

Almost all program logics are for proving the correctness of programs.
Hoare logic is a classic example, whose judgements have the form
$\{P\}\ c\ \{Q\}$ for a program command~$c$ and pre- and postconditions
$P$ and~$Q$. This judgement means that when executed from an initial
state satisfying~$P$ that after command~$c$ finishes, $Q$ is guaranteed to
hold. In this sense postcondition~$Q$ \emph{over-approximates} the final
states that command~$c$ can reach from an initial $P$-state.
Recently, interest has emerged in program logics for proving \emph{incorrectness}~\cite{OHearn_19},
i.e., for diagnosing bugs in programs with a true-positives guarantee.
Such logics inherit the \emph{under-approximate} structure of Reverse Hoare Logic~\cite{deVries_Koutavas_11}.
Their judgements~$[P]\ c\ [Q]$ mean that for all final states~$t$ satisfying
$Q$, there exists an initial $P$-state from which $Q$ can execute to
terminate in state~$t$.
Thus $Q$ under-approximates the final
states that command~$c$ can reach from an initial $P$-state.

\label{sec:nondet}
While the two approaches are roughly equivalent for deterministic programs,
under-approximate reasoning is necessary to accurately diagnose vulnerabilities
in \emph{nondeterministic} programs, including those that 
allocate memory or interact with an outside environment or user.
Incorrectness Separation Logic~\cite{raadlocal,le2022finding} (ISL) is
such an under-approximate logic, which has proved especially useful
for automatic memory-safety bug detection because program analysis in the
logic can be carried out automatically
via bi-abduction based symbolic execution~\cite{calcagno2009compositional,raadlocal}, and supports compositional and incremental
program analysis~\cite{le2022finding}. 

All such under-approximate
logics to-date, however, reason only about individual program executions.
They can therefore detect only those bugs that can be observed in this way,
like assertion failures
(as in Incorrectness Logic~\cite{OHearn_19}) or memory-safety errors
like null-pointer dereferences and use-after-free errors (as in
Incorrectness Separation Logic~\cite{raadlocal}). 
Yet, vulnerabilities come in many kinds, beyond memory-safety issues.
In this paper we focus on the \emph{automatic detection of information leakage vulnerabilities}.
These are especially interesting as they are very common and can be devastating.
But since information leakage is semantically expressed as a
hyperproperty~\cite{Clarkson_Schneider_10},
which compares \emph{pairs} of program executions,
it is out of scope for the existing under-approximative logics.

Can we design an under-approximate logic for reasoning about such
vulnerabilities which inherits the nice property that all defects which are flagged
are true positives?
If so, can analysis using this logic be automated to
produce a compositional vulnerability analysis method?

\textbf{Contribution:}
We answer both of these questions in the affirmative.
In this paper, we present Insecurity Separation Logic (\InsecSL, \cref{sec:insecsl}),
an under-approximate separation
logic for diagnosing information leakage and memory-safety vulnerabilities.
\InsecSL reasons about pairs of program executions but purposefully
closely resembles the (single execution) logic ISL~\cite{raadlocal}.
We show in \cref{sec:symex} how reasoning in \InsecSL can be automated via bi-abduction based
symbolic execution by formalising and proving that the same symbolic
execution procedure as is used for ISL is also sound for \InsecSL.
We demonstrate the practicality of our ideas by implementing them
 in two different tools (\cref{sec:impl}), including an extension of the
Infer tool in which we adapt Infer's ISL implementation to diagnose
information leakage vulnerabilities via \InsecSL. We evaluate our
implementations (\cref{sec:eval}) by applying them to a range of case studies.
Soundness theorems (namely \cref{thm:soundness} for \InsecSL and \cref{thm:symex} for symbolic execution respectively) have been mechanised in Isabelle/HOL. All artifacts
are available online\ifExtended\else\ with the extended version of this paper\fi: \url{https://covern.org/insecurity.html}.

\section{Motivation}\label{sec:motivation}


We use the program in \cref{fig:auction} to both motivate and explain our
approach. This program implements the core of a simple \emph{sealed-bid}
auction server. 
In a sealed-bid auction, all information
about bids must be kept secret until after the auction is finished, at which
point only the winning bid is announced. 

\begin{figure}[t]
  \begin{multicols}{2}
  \begin{lstlisting}[style=customc]
struct bid_t { int id; int qt; };

void run_auction() {
  struct bid_t highest = /* init */;
  while (/* still going */) {
    struct bid_t bid;
    get_bid(&bid);
    update_max(&highest, &bid);
  }
  announce_winner(&highest);
}

void update_max(struct bid_t *a,
                struct bid_t *b)
{
  /* branching on secrets: */
  if (b->qt > a->qt) {
    a->id = b->id;
    a->qt = b->qt;
    /* potentially slow: */
    log_current_max(a->id, a->qt);
  }
}
  \end{lstlisting}
  \end{multicols}
 \vspace*{-5mm}
    \caption{The core of a sealed-bid auction server, adapted from a case-study in SecC:
             \url{https://bitbucket.org/covern/secc/src/master/examples/case-studies/auction.c}.\label{fig:auction}}
\end{figure}

\newcommand{\runauction}{\texttt{run\_auction()}\xspace}
\newcommand{\getbid}{\texttt{get\_bid()}\xspace}
\newcommand{\idqtmax}{\texttt{update\_max()}\xspace}
\newcommand{\announcewinner}{\texttt{announce\_winner()}\xspace}
\newcommand{\bidid}{\mathit{id}}
\newcommand{\bidqt}{\mathit{qt}}
\newcommand{\vars}{\texttt{highest}\xspace}
\newcommand{\vard}{\texttt{status}\xspace}
\newcommand{\bid}{\texttt{bid}\xspace}

Bids in this auction are pairs of \texttt{int}s: $(\bidid,\bidqt)$ where
$\bidid$ identifies the bidder who submitted the bid, and
$\bidqt$ is the amount (or \emph{quote}) submitted in the bid.
The C struct
type \texttt{bid\_t} pairs these two values together. 
The top-level function \runauction maintains
the current maximum bid~\vars,
and a temporary \bid used to store newly submitted bids, which are received via
the \getbid function.
Each new bid is then compared to the current highest one
using the function \idqtmax,
which potentially updates the current highest bid and
persists a record about this fact via \texttt{log\_current\_max}.
%
Note that \getbid is
inherently nondeterministic: It may return arbitrary values, since it is
the interface between the program and its environment.
This puts it outside the scope of Relational Symbolic Execution~\cite{farina2019relational}
as implemented in tools like Binsec/Rel~\cite{daniel2020binsec}.

Unfortunately, \idqtmax is insecure.
As it updates the maximum bid only when the newly submitted bid is larger than
the current maximum, its \emph{timing} depends on whether the branch is taken or not.
This timing leak can be exploited
by auction participants to game the auction.
In particular if \texttt{log\_current\_max} incurs a notable delay%
---writing to disk or even network storage synchronously may be slow---%
they might be capable to infer whether
the bid they have submitted is greater than the current maximum or not.
Moreover, the call to \announcewinner is potentially insecure
under the premise that we only want to disclose the winning bid.
If \vars has not been computed correctly,
then we may accidentally reveal sensitive information about another bid.

\textbf{Challenge:} 
The question of whether a \emph{potential} information leak in a program
becomes critical therefore strongly depends on the context in which functions like \idqtmax and \announcewinner are called.
A compositional underapproximative analysis like that of \InsecSL
must therefore be capable of tracking such relationships \emph{precisely}
to be sound, i.e., to avoid false positives.

\newcommand{\bqt}{\mathit{bqt}}
\newcommand{\aqt}{\mathit{aqt}}
As an example, the security-related summary inferred for \idqtmax, shown below,
expresses that each potentially insecure final state as marked by \textcolor{inseccolor}{$\mathit{insec}$}
is guaranteed to be reachable under the sufficient presumption
that parameters \texttt{a} and \texttt{b} are valid pointers.
Assertion $\Insec{(\bqt > \aqt)}{\ell}$ denotes that this insecurity occurs
if within a given calling context the outcome of the conditional $\bqt > \aqt$
is \emph{not already known to the attacker} of security level~$\ell$ (cf. \cref{sec:attacker} and \cref{sec:insecsl}).
\[
\begin{array}{c}
  \textcolor{blue}{[\pto{\texttt{\&b->qt}}{\bqt} \star \pto{\texttt{\&a->qt}}{\aqt}]} \\ \textcolor{black}{\texttt{update\_max(a,b)}} \\ \textcolor{inseccolor}{[{\it insec\!:}\ \Insec{(\bqt > \aqt)}{\ell} \star \pto{\texttt{\&b->qt}}{\bqt} \star \pto{\texttt{\&a->qt}}{\aqt}]}
  \end{array}
\]
Note that this summary
is beyond the scope of type systems like~\cite{Sabelfeld_Myers_03}
which just capture whether information flow happens or not,
but which fail to adequately reason about logical conditions like $\Insec{(\bqt > \aqt)}{\ell}$.

\section{Attacker Model}\label{sec:attacker}

  We imagine that the execution of
  the program in question is being observed by an \emph{attacker}, who
  has certain observational powers and initial knowledge and is trying to
  deduce secret information that the program is trying to protect. An
  information leak occurs if the attacker can deduce some secret information
  that they did not already know initially before the program was executed.
  
  As standard, the attacker is assumed to know
  the program being executed and
  certain initial values in memory as specified by assertions characterising pre-states.
  The program may perform inputs and outputs during its execution and the attacker is assumed to be able to observe some of these.
  All other information is considered \emph{secret}, and information flow security requires
  that the attacker can never learn any new information above that which they
  were assumed to know initially.
  As usual, we therefore define what an attacker can observe with the help of a security lattice
  comprised of labels~$\ell$ which are comparable by a binary relation~$\sqsubseteq$
  with $\low$ and $\high$ being the least resp. greatest elements,
  modeling public and fully sensitive information, respectively.
  A channel at level~$\ell'$ is observable by an $\ell$-attacker
  if $\ell' \sqsubseteq \ell$, e.g., the $\low$ channel is observable publicly.

  As motivated in \cref{sec:motivation}, the security property for \InsecSL is \emph{timing-sensitive}.
  This means that the attacker can not just observe inputs and outputs on
  certain channels, but also at what times they occur.
  As is typical, time is measured in terms of the number of small-steps
  of execution in the language's small-step operational semantics.
  Following the standard \emph{program counter (PC) security model}~\cite{molnar2005program}, the security property targeted by \InsecSL assumes an
  attacker who is able to observe at each point in time (i.e.\ after each
  small-step of the semantics) the program code that is running.
  This implies that e.g.\
  when executing an if-conditions~$\ITE{e}{c_1}{c_2}$ where $c_1 \not= c_2$,
  that the attacker
  can infer some information about~$e$ (namely whether it evaluated to true
  or not), since they will be able to tell in the subsequent execution
  step whether~$c_1$ or~$c_2$ is being executed.  A similar argument applies
  to while-loops.
  While not as strong as \emph{constant-time security}~\cite{barthe2019formal},
  \InsecSL can be easily extended to cover the stronger attacker
  model of constant-time security if desired (see \ifExtended\cref{sec:ct}\else the extended version of this paper~\cite{Extended}\fi). 

%
%
  We emphasize that the choice of this attacker model is a trade-off:
  under this attacker model it is not possible to verify programs
  that have if/while conditions that depend on secrets, even if leakage
  from such conditions is considered acceptable in certain situations.
On the other hand, a PC-security security guarantee
requires one to consider only ``matched'' executions,
as exploited by \SecCSL~\cite{Ernst_Murray_19} and also by \InsecSL,
which drastically simplifies the logic and its automation
in comparison to product constructions like~\cite{Eilers2018}.

\section{Insecurity Separation Logic (\InsecSL)}\label{sec:insecsl}

Insecurity Separation Logic (\InsecSL) is the relational analogue of
ISL~\cite{raadlocal} and the underapproximative
dual to Security (Concurrent) Separation Logic (\SecCSL)~\cite{Ernst_Murray_19}.
Judgements in \InsecSL are written as
\begin{align}
\iprove{\ell}{P}{c}{Q}
    \label{eq:judgement}
\end{align}
where relational assertions~$P$ characterizes the pre-states
(``presumption'') and~$Q$ characterize
reachable final states (``result''),
$\ell$ is a security level,
$c$ is a program command,
and $\epsilon$ is a status flag that indicates whether the command has
terminated normally ($\epsilon = \color{okcolor}\mathit{ok}$),
whether a runtime error has occurred ($\epsilon = \color{errcolor}\mathit{err}(L)$),
or whether an insecurity has been detected ($\epsilon = \color{inseccolor}\mathit{insec}(L)$).
The latter two track a program location $L$ that points to the
cause of the defect.

The capability to precisely characterise insecurity for nondeterministic programs
is what distinguishes \InsecSL from prior logics.
%
As an example, \InsecSL allows us to derive that the output of the value of an expression~$e$ to a channel of security level $\ell'$ can be potentially witnessed as insecure
without further presumptions
in any (pair of final) state(s) in which~$e$ is secret wrt.~$\ell'$, written $\Insec{e}{\ell'}$, under the assumption of an $\ell$-attacker (which implies $\ell' \sqsubseteq \ell$):
\vspace*{-3mm} 
\begin{align}
  \infer{\ }{\iproveINSEC{\ell}{\Emp}{\Label{L} \Output{\ell'}{e}}{L}{\Insec{e}{\ell'}}}\textsc{OutInsec}
  \label{eq:rule-out-insec}
\end{align}
Judgement~\eqref{eq:judgement} is defined relative to
a relational semantics of assertions like $\Insec{e}{\ell'}$ and $\Emp$,
written $(s,h)\ (s',h') \models_\ell P$ where $s,s'$ are stores
(mappings from variables to values)
and $h,h'$ are heaps (mappings from addresses to values),
and a small-step program semantics
$\sems{k_1}{\sigma}{k_2}$
where configurations $k$ are either
a running program $k_1,k_2 = \Run{c}{s}{h}$,
a terminated execution $k_2 = \Stop{s}{h}$
or a program error $k_2 = \Abort{s}{h}$,
where the latter two correspond to a final status~$\epsilon$
of~$\color{okcolor}\mathit{ok}$
and~$\color{errcolor}\mathit{err}(L)$, respectively.

As a hyperproperty, security
cannot be defined solely by looking at the final state
of a single execution,
comprised of the store $s$ and heap $h$ in $\Stop{s}{h}$ configurations.
Instead, we have to compare what is \emph{observable} between
possible pairs of executions.
To capture this notion, execution steps additionally keep track
of relevant events as a schedule~$\sigma$,
which records for example
input events~$\In{\ell'}{v}$ and outputs events~$\Out{\ell'}{v}$
to track a value~$v$ together with the security level $\ell'$
of the respective communication channel.
The key issue for defining a security logic like \InsecSL (and also \SecCSL)
and proving soundness of rules like~\eqref{eq:rule-out-insec}
is therefore to connect the three ingredients,
namely the judgements~\eqref{eq:judgement},
observations~$\sigma$, and the assertions~$P$,~$Q$ encountered throughout a derivation.
It is based on the following semantic notion:
\begin{definition}[Execution Witness]
    \label{def:witness}
Presumption~$P$ and result~$Q$ witness
an execution of program~$c$ against the $\ell$-level attacker
and a given status~$\epsilon$ when for all
final states~$s$, $h$, $s'$, $h'$ such that
$(s,h)\ (s',h') \vDash_\ell Q$,
there exist initial states $s_0$, $h_0$, $s_0'$, $h_0'$,
and $\sigma$, $\sigma'$, $k$, $k'$
such that $(s_0,h_0)\ (s_0',h_0') \vDash_\ell P$
and
$\semsstar{\Run{c}{s_0}{h_0}}{\sigma}{k}$ and
$\semsstar{\Run{c}{s_0'}{h_0'}}{\sigma'}{k'}$,
where $\sigma$ and~$\sigma'$
have equal lengths and are \emph{input-equivalent}
for the $\ell$-level attacker  (\cref{def:io-equiv}),
and the final store and heap of~$k$ are respectively~$s$ and $h$ and likewise
for $k'$, $s'$ and~$h'$.
Moreover,
\begin{itemize}
\item[] If $\epsilon = \color{okcolor}\mathit{ok}$
      resp. $\epsilon = \color{errcolor}\mathit{err}(L)$ then
    \begin{itemize}
    \item $\sigma$ and~$\sigma'$ are \emph{output-equivalent} to the
      $\ell$-level attacker (\cref{def:io-equiv}),
    \item and $k$ and $k'$ must both be $\kw{stop}$ped resp. $\kw{abort}$ed.
    \end{itemize}
\item[] If $\epsilon = \color{inseccolor}\mathit{insec}(L)$ then
    \begin{itemize}
    \item either~$\sigma$ and~$\sigma'$ are \emph{not} output-equivalent to the
      $\ell$-level attacker,
    \item or $k$ and~$k'$ both denote $\kw{run}$ning configurations with
      different commands.
    \end{itemize}
\end{itemize}
\end{definition}
Witnessing an insecure behaviour therefore violates the standard
security condition of program counter (PC) security~\cite{molnar2005program}.
Also note that the conditions are mutually exclusive, i.e.,
an execution witness can uniquely be classified into an
ok behavior, an erroneous behavior, or an insecure one.

\begin{theorem}[True Positives]\label{thm:soundness}
\InsecSL guarantees that if $\iprove{\ell}{P}{c}{Q}$ is derivable via the rules,
shown in \cref{fig:insecsl-rules},
then there is an execution witness for~$P$, $Q$, $c$, and $\epsilon$
wrt. an $\ell$-attacker, according to \cref{def:witness}.
\end{theorem}



\subsubsection{Assertions.}
\InsecSL assertions are \emph{relational}~\cite{Yang07,Ernst_Murray_19};
pure assertions~$\rho$
and spatial assertions~$P$, $Q$ are defined according to the following grammar:
\begin{align*}
\rho & ~::=~ e \mid \rho \Longrightarrow \rho \mid \Sec{e}{e_\ell} \mid \Insec{e}{e_\ell}
\\
P,Q & ~::=~ \Emp \mid \rho \mid \pto{e}{e'} \mid \pinvalid{e}{} \mid P \star Q \mid \exists x. \ P \mid P \Longrightarrow Q
\end{align*}
where~$e$ ranges over pure
expressions,
including boolean propositions (first case of~$\rho$),
similarly, $e_\ell$ ranges over pure expression
that denote security labels of some designated data type
that models the security lattice and includes constants $\low$ and $\high$
but is not further specified here.

Semantically, assertions are evaluated
over \emph{pairs} of states, written $s\ s' \models_\ell \rho$ and $(s,h)\ (s',h) \models_\ell P$ for stores~$s,s'$ and heaps~$h,h'$,
where the unprimed resp. primed states come from the two executions
being compared.
Stores are mappings from variable names to values as usual,
whereas heaps~$h \colon \Val \rightharpoonup \Val \union \{\bot\}$
are partial functions that include an additional~$\bot$ element as in ISL,
where $p \in \dom{h}$ and $h(p) = \bot$ means that pointer~$p$
is definitely invalid in contrast to $p \notin \dom{h}$,
which means we do not currently have access resp. own~$p$.

The key definitions are as follows (see \ifExtended\cref{fig:rel-sems} \else  \cite{Extended} \fi for the full list):
  \begin{align}
  s\ s' \vDash_{\ell} e           &\iff \eval{e}{s} = \kw{true} \land \eval{e}{s'} = \kw{true} 
        \label{eq:sem-pure} \\[4pt]
  s\ s' \vDash_{\ell} \Sec{e}{e_\ell} &\iff \eval{e_\ell}{s} \sqsubseteq \ell \land \eval{e_\ell}{s'} \sqsubseteq \ell \implies \eval{e}{s} = \eval{e}{s'}
        \label{eq:sem-sec} \\[4pt]
  s\ s' \vDash_{\ell} \Insec{e}{e_\ell} &\iff \eval{e_\ell}{s} \sqsubseteq \ell \land \eval{e_\ell}{s} \sqsubseteq \ell \land \eval{e}{s} \not= \eval{e}{s'}
        \label{eq:sem-insec}
  \end{align}
  where
  we define $(s,h)\ (s',h') \models \rho$ iff $s\ s \models \rho$ and $h = h' = \emptyset$,
  and $\eval{e}{s}$ denotes the evaluation of pure expression~$e$ in store~$s$,
  and $\sqsubseteq$ is the partial order between security labels.
Conditions $\eval{e_\ell}{s} \sqsubseteq \ell$ and $\eval{e_\ell}{s'} \sqsubseteq \ell$
therefore mean that~$e_\ell$ denotes a security label that is relevant wrt.
the ``current'' $\ell$-attacker from $\models_\ell$ resp. \eqref{eq:judgement}.

We can assert a pure boolean expression~$e$ if it is known to hold in both states~$s$ and~$s'$ \eqref{eq:sem-pure}.
Assertion $\Sec{e}{e_\ell}$ denotes \emph{agreement}
of value~$e$ with respect to the security label denoted by~$e_\ell$,
i.e., the value of $e$ is the same in both~$s$ and~$s'$ \eqref{eq:sem-sec}.
It coincides with $\mathbb{A}\,e$ of \cite{banerjee2008expressive} for~$e_\ell = \low$
but just as in SecCSL~\cite{Ernst_Murray_19},
$e_\ell$ can be a more complex expression, not just a constant.
It expresses that an $e_\ell$-attacker knows the value of~$e$,
specifically $\Sec{e}{\low}$ means that~$e$ is public.
Dually, \emph{disagreement} $\Insec{e}{e_\ell}$ formalises that
an attacker who can observe level $e_\ell$ has some uncertainty about~$e$ \eqref{eq:sem-insec}.
Semantically, $s,s' \models_\ell \Insec{e}{e_\ell}$ requires that it is possible
for the expression~$e$ to take two \emph{different} values
in the two stores~$s$ and~$s'$ being compared.
Therefore, leaking the value of~$e$ to an $e_\ell$-visible output channel
is insecure because the attacker can learn
whether the system is actually in state~$s$ or in~$s'$ by observing the value of~$e$.

The second feature for bug-detection is
the assertion $\pinvalid{e}$ from ISL~\cite{raadlocal},
which expresses that~$e$ is known to be an invalid pointer,
so that dereferencing~$e$ is necessarily incorrect.
This is dual to the standard points-to assertion $\pto{e}{e'}$
which states that memory location~$e$ is valid and contains value~$e'$.

We point out that relational implication $\implies$ is distinct
from pure implication at the level of expressions (not shown here).
All other connectives intuitively mean the same as in a non-relational setting,
e.g.,
$\Emp$ denotes an empty heap and $P * Q$ asserts $P$ and $Q$ on two disjoint parts of the heap,
but of course technically these have to be lifted to the relational setting semantically.


\subsubsection{Commands and Semantics.}
  Commands~$c$ in the language are as follows, where $e$ is a pure expression that can mention program variables~$x$:
  \[
  \begin{array}{l@{\,}l}
    c ~::=~ & \Skip \mid \Assign{x}{e} \mid \Load{x}{e} \mid \Store{e}{e'} \mid \Alloc{x}{e} \mid    \Free{e} \mid  \\
    & \Label{L} c \mid \Seq{c_1}{c_2} \mid \ITE{e}{c_1}{c_2} \mid \While{e}{c}  \mid \\
    & \Output{e}{e'} \mid \Input{x}{e}
    \end{array}
  \]
  Here $[e]$ denotes dereferencing pointer~$e$ and e.g.\ in C would be written \texttt{*e}.
  As in ISL~\cite{raadlocal}, commands in \InsecSL carry an optional
  label~$L$ that is used for error-reporting, written $\Label{L} c$.
  Most commands are standard, except~$\Input{x}{e_\ell}$ and $\Output{e_\ell}{e}$.
  Command $\Input{x}{e_\ell}$ means input a value from the channel denoted by~$e_\ell$ and assign the inputted value to the variable~$x$;
  command $\Output{e_\ell}{e}$ means
  to output the value denoted by the expression~$e$ on the output channel
  denoted by the expression~$e_\ell$. 

  The language of \InsecSL is given a small-step semantics~$\sems{k_1}{\sigma}{k_2}$,
  allowing judgements to talk about partial executions ending in \kw{run}ning
  non-final states
  (cf. {\color{inseccolor}$insec(L)$} case in \cref{def:witness}).
  Importantly, this semantics records the values and security labels of input and output commands as part of schedule~$\sigma$,
  which is necessary to state the formal security properties used for \InsecSL's
  soundness result in \cref{thm:soundness} via \cref{def:io-equiv} below.

  The schedule is a list of events
$e ::= \tau \mid \In{\ell}{v} \mid \Out{\ell}{v} \mid \Allocate{v}$
  for security level~$\ell$ and value~$v \in \Val$.
  Event $\tau$ represents a single, non-$\kw{input}$,
  non-$\kw{output}$,
  non-$\kw{alloc}$ step of computation, i.e., $\tau$ steps are not critical for security.
  Event $\In{\ell}{v}$ records that value~$v$
  was input at security level~$\ell$ and
  $\Out{\ell}{v}$ records that
  value~$v$ was output at security level (i.e.\ on the output channel)~$\ell$, while $\Allocate{v}$
  records that address~$v$ was dynamically allocated.
  It is simply included
  as a convenience to ensure that all non-determinism can be resolved by the schedule~$\sigma$.
  Some key rules are shown below, the full listing is in \ifExtended\cref{fig:sems}\else\cite{Extended}\fi.
  \begin{mathpar}
  \infer{a = \eval{p}{s} \and h(a) = v}{\sems{\Run{\Load{x}{p}}{s}{h}}{\singlist{\tau}}{\Stop{\funupd{s}{x}{v}}{h}}}

  \infer{a = \eval{p}{s} \and a \not\in \dom{h} \lor h(a) = \bot}{\sems{\Run{\Load{x}{p}}{s}{h}}{\singlist{\tau}}{\Abort{s}{h}}}

  \infer{}{\sems{\Run{\Input{x}{e_\ell}}{s}{h}}{\singlist{\In{\eval{e_\ell}{s}}{v}}}{\Stop{\funupd{s}{x}{v}}{h}}}

  \infer{}{\sems{\Run{\Output{e_\ell}{e}}{s}{h}}{\Out{\eval{e_\ell}{s}}{\eval{e}{s}}}{\Stop}{s}{h}}
  \end{mathpar}
%

  The first rule shows a load via pointer expression~$p$ from a valid address~$a$,
  the corresponding value in the heap is then assigned to variable~$x$ in the updated store~$s(x:=v)$.
  Notice that we can observe memory errors in this semantics directly by
  transitions to $\Abort{s}{h}$ configurations,
  as it is for example when the pointer expression~$p$ instead evaluates to
  an unknown address~$a \notin \dom{h}$
  or one that is definitely not allocated~$h(a) = \bot$ (second rule).
  Reading from an input channel returns a non-deterministic value~$v$
  that is assigned to~$x$ in the successor state.
  However, information
  leakage can only be observed by comparing pairs of executions
  in terms of their schedules (cf. \cref{def:witness}).

  As an example, $\Output{e_\ell}{e}$ with $\eval{e_\ell}{s} = \ell'$
  and $\eval{e_\ell}{s'} = \ell'$ in a pair of executions with stores~$s$ and~$s'$
  respectively,
  will expose two schedules $\sigma = \singlist{\Out{\ell'}{v}}$
  and $\sigma' = \singlist{\Out{\ell'}{v'}}$,
  where $v = \eval{e}{s}$ and $v'= \eval{e}{s'}$ are the values
  that are output over the channel in the two runs.
  If $\ell' \sqsubseteq \ell$, i.e., the channel is visible to the attacker,
  then an information leak occurs if $v \neq v'$ and we have an execution witness
  according to \cref{def:witness} and \cref{eq:sem-insec} for result $\color{inseccolor}[\mathit{insec}\!: \Insec{e}{\ell'}]$.
  Input- and output-equivalence
  which \cref{def:witness} relies on is therefore as follows:
  \begin{definition}[Input and Output Equivalence]
    \label{def:io-equiv}
  Two schedules are
  are \emph{input} resp. \emph{output equivalent} for the $\ell$-level attacker when all
  inputs resp. outputs observable to that attacker are identical in each,
  i.e., after projecting the schedules to those input resp. output events,
  $\In{\ell'}{v}$ or $\Out{\ell'}{v}$ for which $\ell' \sqsubseteq \ell$.
  \end{definition}

\subsubsection{Proof Rules and Soundness.}
The proof rules of \InsecSL are in \cref{fig:insecsl-rules}.
Rules analog to those of ISL~\cite{raadlocal} are included,
those rules that mention value classification (e.g. in \textsc{Input})
and those with \textcolor{inseccolor}{insec} result are specific to \InsecSL.

%
%
%

\begin{figure}[p]
\scalebox{0.83}{
  \begin{minipage}{1.2\textwidth}
    \begin{mathpar}
      \infer{\ }{\iproveOK{\ell}{x=x'}{\Input{x}{e}}{\Sec{x}{e[x'/x]}}}\textsc{Input}

  \infer{\ }{\iproveOK{\ell}{x=x'}{\Assign{x}{e}}{x = e[x'/x]}}\textsc{Assign}

  \infer{\ }{\iproveOK{\ell}{P}{\Skip}{P}}\textsc{Skip}

  \infer{\ }{\iproveOK{\ell}{x=x' \star \pto{p}{e}}{\Load{x}{p}}{x = e[x'/x] \star \pto{p}{e[x'/x]}}}\textsc{LoadOK}
 
   \infer{\ }{\iproveOK{\ell}{\pto{p}{e}}{\Store{p}{e'}}{\pto{p}{e'}}}\textsc{StoreOK}
    
   \infer{\ }{\iproveERR{\ell}{\pinvalid{p}}{\Label{L} \Load{x}{p}}{L}{\pinvalid{p}}}\textsc{LoadErr}

   \infer{\ }{\iproveERR{\ell}{\pinvalid{p}}{\Label{L} \Store{p}{e}}{L}{\pinvalid{p}}}\textsc{StoreErr}

   \infer{\ }{\iproveOK{\ell}{\Emp}{\Alloc{x}{e}}{\pto{x}{e}}}\textsc{Alloc1}
   
  \infer{\ }{\iproveOK{\ell}{\pinvalid{p}}{\Alloc{x}{e}}{x = p \star \pto{p}{e}}}\textsc{Alloc2}

  \infer{\ }{\iproveOK{\ell}{\pto{p}{e}}{\Free{p}}{\pinvalid{p}}}\textsc{FreeOK}

  \infer{\ }{\iproveERR{\ell}{\pinvalid{p}}{\Label{L} \Free{p}}{L}{\pinvalid{p}}}\textsc{FreeErr}

  \infer{\ }{\iproveOK{\ell}{\Emp}{\Output{\ell'}{e}}{\Sec{e}{\ell'}}}\textsc{OutOK}

  \infer{\ }{\iproveINSEC{\ell}{\Emp}{\Label{L} \Output{\ell'}{e}}{L}{\Insec{e}{\ell'}}}\textsc{OutInsec} \\

  \infer{\iprove{\ell}{b \star P}{c_1}{Q}}{\iprove{\ell}{P}{\ITE{b}{c_1}{c_2}}{Q}}\textsc{IfTrue}

  \infer{\iprove{\ell}{\lnot b \star P}{c_2}{Q}}{\iprove{\ell}{P}{\ITE{b}{c_1}{c_2}}{Q}}\textsc{IfFalse}  

  \infer{c = \Label{L} \ITE{b}{c_1}{c_2} \and c_1 \not= c_2}{\iproveINSEC{\ell}{\Insec{(b = \true)}{\ell}\star F}{c}{L}{\Insec{(b = \true)}{\ell}\star F}}\textsc{IfInsec}

  \infer{\iprove{\ell}{b \star P}{\Seq{c}{\While{b}{c}}}{Q}}{\iprove{\ell}{P}{\While{b}{c}}{Q}}\textsc{WhileTrue}

  \infer{\ }{\iproveOK{\ell}{\lnot b\star F}{\While{b}{c}}{\lnot b\star F}}\textsc{WhileFalse}

   \infer{\ }{\iproveINSEC{\ell}{\Insec{(b = \true)}{\ell}\star F}{\Label{L} \While{b}{c}}{L}{\Insec{(b = \true)}{\ell}\star F}}\textsc{WhileInsec}

  \infer{\iproveOK{\ell}{P}{c_1}{Q} \and \iprove{\ell}{Q}{c_2}{R}}{\iprove{\ell}{P}{\Seq{c_1}{c_2}}{R}}\textsc{SeqOK}
 
  \infer{\iproveERR{\ell}{P}{c_1}{L}{Q}}{\iproveERR{\ell}{P}{\Seq{c_1}{c_2}}{L}{Q}}\textsc{SeqErr}
  
  \infer{\iproveINSEC{\ell}{P}{c_1}{L}{Q}}{\iproveINSEC{\ell}{P}{\Seq{c_1}{c_2}}{L}{Q}}\textsc{SeqInsec}

  \infer{\iprove{\ell}{P}{c}{Q} \and \mods{c} \inter \fv{R} = \emptyset}{\iprove{\ell}{P \star R}{c}{Q \star R}}\textsc{Frame}

  \infer{\follows{P'}{\ell}{P} \and \iprove{\ell}{P'}{c}{Q'} \and \follows{Q}{\ell}{Q'}}{\iprove{\ell}{P}{c}{Q}}\textsc{Cons}
 
  \infer{\iprove{\ell}{P_1}{c}{Q_1} \and \iprove{\ell}{P_2}{c}{Q_2}}{\iprove{\ell}{P_1 \lor P_2}{c}{Q_1 \lor Q_2}}\textsc{Disj}
  
  \infer{\iprove{\ell}{P}{c}{Q} \and x \not\in \fv{c}}{\iprove{\ell}{\exists x.\ P\ x}{c}{\exists\ x. \ Q\ x}}\textsc{Ex}  
    \end{mathpar}
    \end{minipage}
}
  \caption{The rules of \InsecSL.\label{fig:insecsl-rules}}
\end{figure}

Rule \textsc{LoadErr} captures the case when loading via pointer~$p$
leads to an error, which is reachable from a presumption $p \pinvalid{}$,
i.e., states in which $p$ is definitely an invalid pointer~\cite{raadlocal}.
It is formulated as a ``small axiom'' as typical for separation logic
which is put into larger context by the standard frame rule (which is valid in our setting).
We remark that sequential composition, too, works as expected.

Rule \textsc{Input} derives that the new value of variable~$x$ in the result
can be classified with respect to~$e_\ell$---auxiliary variable~$x'$ is just a technical
artifact to lift~$e$ over the assignment to~$x$ if~$e$ depends on~$x$.
Input commands can never be insecure,
instead, manifest the domain assumption
that only $e_\ell$-attackers can observe the value that has been stored in~$x$
so that~$x$ is rightly classified by the level denoted by~$e_\ell$.
Soundness of the rule therefore considers whether~$\Sec{x}{e_\ell[x'/x]}$
holds in a given trace, i.e., whether
$\eval{x}{s(x:=v)} = v$ equals $\eval{x}{s'(x:=v')} = v'$
in case $e_\ell$ is $\ell$-visible (via \eqref{eq:sem-sec}),
and if not, this pair of traces can be neglected
as respective schedule-fragments~$\sigma = \singlist{{\In{\eval{e_\ell}{s}}{v}}}$ and~$\sigma' = \singlist{{\In{\eval{e_\ell}{s'}}{v'}}}$
from the small-step semantics are not input equivalent (cf. \cref{def:witness}).

In comparison, there are two rules for the output command,
one for a secure output, \textsc{OutOk}, and one for an insecure output, \textsc{OutInsec} shown in \eqref{eq:rule-out-insec}.
If one wants to prove for a given case study that
the insecure outcome~$\Insec{e}{e_\ell}$ is unreachable,
one can check the result and presumption wrt. a frame assertion~$P$
that captures the path condition of the context in which the output was made,
so that if $P * \Insec{e}{e_\ell}$ is unsatisfiable the result is demonstrated to be unreachable.

Moreover, there are rules that expose branching on secrets as the test of $\textbf{if}$ and $\textbf{while}$ statements,
and rule \textsc{SeqInsec} propagates an insecurity in the first part of a sequential composition similarly to an error.

\section{Symbolic Execution}\label{sec:symex}

\InsecSL's careful design, as a relational logic that resembles the
non-relational ISL, means that its application can be automated via
bi-abduction~\cite{calcagno2009compositional} based symbolic execution method for automatically
deriving \InsecSL judgements.

We formalise the symbolic execution method for ISL, atop \InsecSL, proving
that it yields a sound analysis method for automatically inferring
\InsecSL judgements. Ours is the first such symbolic execution method,
for an under-approximate logic, to enjoy a mechanised proof of soundness.

To define our symbolic execution, it helps to introduce an extra program command
$\Assume{e}$. This command is not a ``real'' command in the sense that it
cannot appear in program text. Instead, it is used to remember, during symbolic execution,
which conditional branches have been followed along the current execution
path. As we will see, our symbolic execution maintains a \emph{trace} that records the
execution path followed so far, in which assume commands~$\Assume{e}$ can appear.
Their semantics is to evaluate the condition~$e$ and, if~$e$ holds
to act as a no-op but otherwise execution gets stuck.

\newcommand{\Trans}[4]{\mathsf{transform}_{\ell}(#1,#2,#3,#4)}
\newcommand{\Backp}[3]{\mathsf{backprop}_{\ell}(#1,#2,#3)}
\newcommand{\Symex}[5]{\mathsf{symex}_{\ell}(#1,\ #2,\ #3,\ #4,\ #5)}
\newcommand{\iproveJ}[4]{\vdash_{#1} \textcolor{blue}{[#2]}\ \textcolor{black}{#3}\ \textcolor{blue}{[#4]}}
\newcommand{\OK}[1]{\textcolor{okcolor}{[{\it ok\!:}\ #1]}}
\newcommand{\INSECP}[2]{\textcolor{inseccolor}{[{\it insec(#1)\!:}\ #2]}}
\newcommand{\ERR}[2]{\textcolor{errcolor}{[{\it err(#1)\!:}\ #2]}}

\newcommand{\Cons}[2]{#1:#2}
Our symbolic execution method stores the path followed so far. Doing so allows
it to provide detailed information to the user when a vulnerability is detected
(e.g.\ to tell precisely along which path the vulnerability arises). Doing so
is also necessary to prove the soundness of our method, as explained later.
The
current path is stored 
as a \emph{trace}, which is a list of
pairs $(c,P)$ where $c$ is a program command and~$P$ an \InsecSL assertion.
For convenience, traces are stored in \emph{reverse} order. Each
element $(c,P)$ is understood to mean that command~$c$ was executed
from symbolic state~$P$, i.e.\ $P$ represents the state before~$c$ was
executed.
We write the empty trace $[]$ (which represents that there has been no
preceding symbolic execution), and the trace whose head is $x$ and whose
tail is $xs$ as $\Cons{x}{xs}$.

When a new spatial assertion~$F$ is inferred to make forward progress in
symbolic execution, it is then \emph{back-propagated} along the trace~$tr$, causing
$F$ to be added into each of the assertions~$P$ in each element~$(c,P)$ of~$F$.
Given an assertion~$F$, back-propagating it over trace~$tr$
produces the transformed trace~$tr'$, and operates in the expected way by
successively appealing to the \textsc{Frame} rule. We define the procedure $\Backp{F}{tr}{tr'}$ for doing this.

\begin{definition}[Backprop] \label{defn:backprop}
	For any assertion $F$, any security level $\ell$, and any traces $tr$ and $tr'$ where each of them is a list of command-assertion pairs,
	 $\Backp{F}{tr}{tr'}$ holds if and only if: $tr=tr'=[] \lor
	(\exists c\ P\ F\ F'\ tr_2\ tr_2'.\ 
	\ tr=\Cons{(c,P)}{tr'}\land tr'=\Cons{(c,P\star F)}{tr_2'}\land \mods{c} \inter \fv{F} = \emptyset \land\Backp{F'}{tr_2}{tr_2'})$
\end{definition}

Symbolic execution is then defined as follows. We define a judgement \linebreak
$\Symex{tr}{JQ}{c}{tr'}{JQ'}$. Here $c$ is a command,
$tr$ and $tr'$ are traces, while
$JQ$ and $JQ'$ are
judgement \emph{post assertions}, i.e.\ have one of the following forms
each for some assertion~$Q$: $\textcolor{okcolor}{\mathit{ok}\!: Q}$,
$\textcolor{errcolor}{\mathit{err}\!: Q}$, or
$\textcolor{inseccolor}{\mathit{insec}\!: Q}$. 
Trace $tr$ and $JQ$ represent the current state of symbolic execution
before command~$c$ is executed, in the sense that $tr$ is the trace
followed up to this point and $JQ$ represents the symbolic state immediately
before $c$ is executed. Executing~$c$ necessarily extends the trace (possibly
also transforming it via back-propagation), yielding an updated trace~$tr'$
and a new post assertion~$JQ'$.
%

The symbolic execution rules are shown in \cref{fig:symex}.
When encountering branching, symbolic
execution will flag insecurity (\textsc{SEIfInsec}) if the branch condition~$b$ is secret
($\Insec{b = \true}{\ell}$); however it can also proceed (e.g. \textsc{SEIfTrue}) by assuming the
branch condition (implicitly assuming it is non-secret). The rule
\textsc{SEOutInsec} detects insecure outputs. Rules for inferring spatial
predicates via bi-abduction follow their counterparts in ISL~\cite{le2022finding}.
%
%
%
%

	\begin{figure}[ht!]
    \scalebox{0.82}{
		\begin{mathpar}
			\infer{\ }{\Symex{tr}{\OK{P}}{\Skip}{\Cons{(\Skip,P)}{tr}}{\OK{P}}}
			\textsc{SESkip}
			
			\infer{\ }
			{\Symex{tr}{\OK{P}}{\Assume{b}}{\Cons{(\Assume{b},P)}{tr}}{\OK{P \star b}}}
			\textsc{SEAsm}
			
			\infer{\ }
			{\Symex{tr}{\OK{P}}{\Output{el}{e}}{\Cons{(\Output{el}{e},P)}{tr}}{\OK{P\star \Sec{e}{el}}}}
			\textsc{SEOut}
			
			\infer{c = (\Output{el}{e})}
			{\Symex{tr}{\OK{P}}{\Label{L}c}{\Cons{(\Label{L}c,P)}{tr}}{\INSECP{L}{P\star \Insec{e}{el}}}}
			\textsc{SEOutInsec}
			
			\infer{x'\notin \fv{P}}
			{\Symex{tr}{\OK{P}}{\Assign{x}{e}}{\Cons{(\Assign{x}{e},P)}{tr}}{\OK{P[x'/x]\star x=e[x'/x]}}}
			\textsc{SEAssign}

			\infer{x'\notin \fv{P}}
			{\Symex{tr}{\OK{P}}{\Input{x}{e}}{\Cons{(\Input{x}{e},P)}{tr}}{\OK{P[x'/x]\star \Sec{x}{e[x'/x]}}}}
			\textsc{SEInput}
			
			\infer{x'\notin \fv{P}}
			{\Symex{tr}{\OK{P}}{\Alloc{x}{e}}{\Cons{(\Alloc{x}{e},P)}{tr}}{\OK{P[x'/x]\star \pto{x}{e[x'/x]}}}}
			\textsc{SEAlloc}
			
			\infer{\Backp{M}{tr}{tr'}\and x'\notin \fv{Frame} \and 
				\follows{\pto{p}{e}\star Frame}{\ell}{P\star M}}
			{\Symex{tr}{\OK{P}}{\Load{x}{p}}{\Cons{(\Load{x}{p},P\star M)}{tr'}}{\OK{x=e[x'/x] \star (\pto{p}{e}\star Frame)[x'/x]}}}
			\textsc{SELoad}
			
			\infer{\Backp{M}{tr}{tr'} \and \follows{\pinvalid{p} \star Frame}{\ell}{P \star M}}
			{\Symex{tr}{\OK{P}}{\Label{L}\Load{x}{p}}{\Cons{(\Label{L}\Load{x}{p},P\star M)}{tr'}}{\ERR{L}{\pinvalid{p}\star Frame}}}
			\textsc{SELoadErr}
			
			\infer{\Backp{M}{tr}{tr'} \and 
			\follows{\pto{p}{e}\star Frame}{\ell}{P\star M}}
			{\Symex{tr}{\OK{P}}{\Store{p}{e'}}{\Cons{(\Store{p}{e'},P\star M)}{tr'}}{\OK{\pto{p}{e'}\star Frame}}}
			\textsc{SEStore}
			
			\infer{\Backp{M}{tr}{tr'} \and \follows{\pinvalid{p}\star Frame}{\ell}{P\star M}}
			{\Symex{tr}{\OK{P}}{\Label{L}\Store{p}{e'}}{\Cons{(\Label{L}\Store{p}{e'},P\star M)}{tr'}}{\ERR{L}{\pinvalid{p}\star Frame}}}
			\textsc{SEStoreErr}
			
			\infer{\Backp{M}{tr}{tr'} \and 
			\follows{\pto{p}{e}\star Frame}{\ell}{P\star M}}
			{\Symex{tr}{\OK{P}}{\Free{p}}{\Cons{(\Free{p},P\star M)}{tr'}}{\OK{\pinvalid{p}\star Frame}}}
			\textsc{SEFree}
			
			\infer{\Backp{M}{tr}{tr'} \and \follows{\pinvalid{p}\star Frame}{\ell}{P\star M}}
			{\Symex{tr}{\OK{P}}{\Label{L}\Free{p}}{\Cons{(\Label{L}\Free{p},P\star M)}{tr'}}{\ERR{L}{\pinvalid{p}\star Frame}}}
			\textsc{SEFreeErr}
                        
		\infer{c = (\ITE{b}{c1}{c2}) \and c1\neq c2}
		{\Symex{tr}{\OK{P}}{\Label{L}c}{\Cons{(\Label{L}c,P\star \Insec{b = \true}{\ell})}{tr}}{\INSECP{L}{P\star \Insec{b = \true}{\ell}}}}
		\textsc{SEIfInsec}
                        
		\infer{\Symex{tr}{\OK{P}}{\Seq{\Assume{b}}{c_1}}{tr'}{Q}}
		{\Symex{tr}{\OK{P}}{\ITE{b}{c_1}{c_2}}{tr'}{Q}}
		\textsc{SEIfTrue}

		\end{mathpar}
    }
	\caption{Symbolic execution rules.\label{fig:symex}}
	\end{figure}

%
\begin{theorem}[Soundness of Symbolic Execution]\label{thm:symex}
	For all commands $c$, security levels~$\ell$, post-assertions $JQ$ and~$JQ'$
	and all traces $tr$,
	produced by symbolic execution,
    i.e., $\Symex{[]}{JQ}{c}{tr}{JQ'}$ holds, we have $tr$ is
        not empty. Furthermore,
	letting $(c,P)$ denote the last element of~$tr$, we have
	$\iproveJ{\ell}{P}{c}{JQ'}$.
\end{theorem}

As mentioned earlier, the trace $tr$ is not merely a user convenience but
a necessary ingredient to prove soundness of the structural rules,
like \textsc{SEIfTrue} above. Soundness of this rule
for instance requires deducing a judgement $\iprove{\ell}{P}{\Seq{c_0}{c'}}{Q}$
given premise $\iprove{\ell}{P}{\Seq{c_0}{c}}{Q}$ and
inductive hypothesis $\forall P\ Q.\ \iprove{\ell}{P}{c}{Q} \implies \iprove{\ell}{P}{c'}{Q}$.
Unfortunately the premise is not strong enough to deduce some intermediate
assertion $R$ for which $\iprove{\ell}{P}{c_0}{R}$ and $\iprove{\ell}{R}{c}{Q}$
as required to instantiate the inductive hypothesis.
Inclusion of trace~$tr$ allows us to express the necessary strengthening of the theorem.
This construction was not necessary for the pen-and-paper
soundness proof of ISL~\cite{raadlocal,le2022finding} because
for any single state there exists an ISL assertion that precisely
describes that state, and hence the existence of the intermediate assertion
$R$ is trivial in ISL. The same is not true for \InsecSL because \InsecSL's
assertions, while resembling unary ones,
are evaluated relationally (cf. \cref{sec:insecsl}).


Our symbolic execution as described can be applied to the body of a function
to infer \InsecSL judgements that describe its internal
behaviour. Such judgements must be transformed
into summaries that describe the function's external behaviour. To do so
we follow the same approach as in ISL~\cite{le2022finding}.
For instance, consider
the trivial function \texttt{void func(int x)\{ x = x + 1; \}} that
uselessly increments its argument~\texttt{x}. Its internal behaviour
is captured by the judgement $\iproveOK{\ell}{\texttt{x} = v}{\texttt{x = x + 1}}{v' = v \star \texttt{x} = v' + 1}$, where the logical variable~$v$ captures
the initial value of \texttt{x}. Transforming this internal judgement into
an external summary (after simplification) yields the summary
  $\iproveOK{\ell}{\Emp}{\texttt{func(x)}}{\Emp}$.

\section{Implementation}\label{sec:impl}

We implemented the symbolic execution procedure for automating the
application of \InsecSL in two tools: \Tool and \InferTool.
\Tool implements the entirety of
\InsecSL via \emph{contextual, top-down} inter-procedural symbolic execution. \InferTool on the other hand is a modification of the existing
non-contextual, \emph{bottom-up} inter-procedural symbolic execution
method for ISL that is implemented in the Pulse-ISL plugin for Infer~\cite{le2022finding}, which we modify to implement a useful subset of the
\InsecSL logic.


\smallskip

\Tool is a proof-of-concept tool, which we built by modifying an existing
verifier for the over-approximate security separation logic \SecCSL~\cite{Ernst_Murray_19}. \Tool implements a top-down inter-procedural
analysis in which individual functions (procedures) are analysed using the
symbolic execution method of \cref{sec:symex} to derive summaries for
their behaviours.

When analysing a function~$f()$ that calls another~$g()$ \Tool attempts to apply
all summaries known about~$g()$. If none of them are applicable (i.e.\ applying
them yields an inconsistent state), \Tool performs a contextual analysis of
$g()$ to compute new summaries applicable at this callsite.
To perform a contextual analysis of callee~$g()$ from
caller~$f()$ we take the current symbolic state~$R$ and filter it to
produce a state~$R'$ that describes only those parts of~$R$ relevant to the
call.
\Tool's present implementation does so using a fixed-point computation that
identifies all pure formulae from~$R$ that mention arguments passed to~$g()$
and values (transitively) related to those arguments by such pure formulae.
It identifies all spatial
assertions in~$R$ that describe parts of the heap reachable from those values,
filtering everything else as irrelevant.

In contrast to Infer~\cite{raadlocal,le2022finding}, \Tool does not unroll loops
to a fixed bound. Instead it controls symbolic execution using two mechanisms.
Firstly, for each program point
it counts the number of paths that have so far passed
through that point during analysis. When that number exceeds a configurable
bound, additional paths are discarded. Additionally it monitors the latency
of symbolically executing each program statement. When this latency gets too
high (exceeds a configurable timeout), the current path is discarded. The former bound is reached only when
unfolding relatively tight loops, while the latter attempts to maintain
reasonable symbolic execution throughput.
When analysing a function \Tool will avoid generating multiple summaries that
report the same problem for a single program point. \Tool reports \emph{unconditional} (aka \emph{manifest}~\cite{le2022finding}) bugs whose presumptions are $\true$.

\Tool encodes all non-spatial formulae to SMT via a relational encoding
which directly encodes their relational semantics \ifExtended(\cref{fig:rel-sems})\else\cite{Extended}\fi.
Doing so necessarily duplicates each variable, meaning that SMT
encodings of formulae are often relatively large. While this can
impede scalability, it ensures that \Tool encodes the
entirety of \InsecSL in a semantically complete way.

\smallskip

\InferTool takes a different design to \Tool, and makes maximum
advantage of the fact that \InsecSL is purposefully designed to be
very similar to ISL~\cite{raadlocal}, allowing its symbolic execution
procedure (\cref{sec:symex}) to very closely resemble that for
ISL also~\cite{le2022finding}.

\InferTool implements a non-trivial fragment of \InsecSL. In this
fragment, there are only two security levels $\ell$: $\low$ (bottom) and
$\high$ (top). The level of the attacker is $\low$. 
Insecurity assertions $\Insec{b}{\low}$ appear only over
boolean expressions~$b$ and mention only the security level $\low$.
Security assertions $\Sec{e}{\low}$ do not appear directly.
Instead, whenever an expression~$e$ is to be treated as $\low$
($\Sec{e}{\low}$), the expression~$e$ is concretised, i.e.\ replaced by
a concrete value (a constant). We refer to this process as
\emph{low concretisation}. Since constants are $\low$ by
definition, concretising $\low$ expressions~$e$ ensures that \InferTool
treats them as $\low$ without having to perform a relational encoding
of the security assertion~$\Sec{e}{\low}$. In our current implementation,
constants for concretisation are not chosen randomly, ensuring
determinism.

Likewise, \InferTool avoids having to perform relational encoding of
insecurity assertions~$\Insec{b}{\low}$ by soundly encoding them as
follows. In particular:
\[
\textbf{sat}\ \Insec{b}{\low} \iff \textbf{sat}\ b \textrm{\ and\ } \textbf{sat}\ \lnot b.
\]
Thus satisfiability of insecurity assertions over boolean conditions~$b$
can be checked via unary (non-relational) satisfiability checking.

With these two techniques, \InferTool automates \InsecSL reasoning
directly within the existing symbolic execution framework for ISL with
minimal modifications, inheriting Infer's highly optimised implementation
and scalability. In this implementation, \InferTool performs
symbolic execution in a bottom-up fashion: each function is analysed in
isolation from all others to produce summaries. Loops are unrolled up to
a fixed bound, making symbolic execution entirely deterministic.

\section{Evaluation}\label{sec:eval}

\newcommand{\auction}{\texttt{auction}\xspace}
\newcommand{\ctselect}{\texttt{ctselect}\xspace}
\newcommand{\ctsort}{\texttt{ctsort}\xspace}
\newcommand{\cttkinner}{\texttt{cttkinner}\xspace}
\newcommand{\haclpolicies}{\texttt{haclpolicies}\xspace}
\newcommand{\hex}{\texttt{hex}\xspace}
\newcommand{\intthirtyone}{\texttt{int31}\xspace}
\newcommand{\kremlib}{\texttt{kremlib}\xspace}
\newcommand{\libsodiumutils}{\texttt{libsodiumutils}\xspace}
\newcommand{\opensslutil}{\texttt{opensslutil}\xspace}
\newcommand{\oram}{\texttt{oram1}\xspace}
\newcommand{\sslcbcrem}{\texttt{ssl3cbcrem}\xspace}
\newcommand{\tlslucky}{\texttt{tls1lucky13}\xspace}
\newcommand{\tlspatched}{\texttt{tls1patched}\xspace}

\begin{table}[t]
  \caption{\label{tbl:results}Tool evaluation results. For each sample we record its size in (SLOC) and the number of \emph{top-level} functions analysed (\# funs). The third column (sec?) indicates whether the sample had no security vulnerabilities known a-priori.
  Analysis time of \Tool for each sample is reported in seconds.
  \InferTool\ analysed each sample in less than one second.
  The unique top-level bugs reported we break down into memory safety errors (\# err, for \Tool)
  and information leaks (\# insec, for both).
  $(\dagger)$ indicates samples that were analysed in \InferTool with a manually set loop unrolling bound.
  $(\ddagger)$ indicates samples that were analysed in \Tool with an increased symbolic execution pruning and SMT timeout of 600 seconds. }

  \setlength{\tabcolsep}{4pt}
  \begin{tabular}{lrrrrrrr}
  \toprule
    &&&& \multicolumn{3}{c}{\Tool}
       & \multicolumn{1}{c}{\InferTool} \\
    \cmidrule(l){5-7}
    \cmidrule(l){8-8}
      Sample
    & SLOC
    & \#\,funs
    & sec?
    & time (s)
    & \# err
    & \# insec
    & \# insec
    \\
    \cmidrule{1-1}
    \cmidrule(l){2-2}
    \cmidrule(l){3-3}
    \cmidrule(l){4-4}
    \cmidrule(l){5-5}
    \cmidrule(l){6-6}
    \cmidrule(l){7-7}
    \cmidrule(l){8-8}
    \auction $(\dagger)$       &  172 &  1  & \xmark & 195 & 0 & 1 & 1 \\
    \ctselect                  &   27 &  5  & \xmark &   1 & 0 & 1 & 1 \\
    \ctsort                    &   57 &  3  & \xmark &   5 & 0 & 2 & \bf 7 \\
    \cttkinner                 &   77 &  3  & \cmark &   5 & 0 & 0 & 0 \\
    \haclpolicies              &   34 &  1  & \cmark &  50 & 0 & 0 & 0 \\
    \hex                       &  178 &  2  & \cmark &  80 & 0 & 1 & 1 \\
    \intthirtyone $(\ddagger)$ & 1923 & 60  & \cmark & 708 & 1 & 2 & - \\
    \kremlib                   &   68 & 10  & \cmark &   2 & 0 & 0 & 0 \\
    \libsodiumutils            &  115 &  3  & \cmark & 380 & 0 & 1 & 1 \\
    \opensslutil               &   84 &  7  & \cmark &   1 & 0 & 0 & 0 \\
    \oram                      &  167 &  4  & \cmark &  27 & 0 & 1 & 1 \\
    \sslcbcrem                 &  111 &  1  & \cmark &  10 & 0 & 0 & 0 \\
    \tlslucky                  &  122 &  1  & \xmark & 119 & 1 & 4 & \bf 6 \\
    \tlspatched                &  229 &  1  & \cmark & 192 & 2 & 2 & 0 \\
    \bottomrule
  \end{tabular}
\end{table}

We evaluate both \Tool and \InferTool on the programs,
listed in \cref{tbl:results}. The \auction sample is the synthetic
auction case study from \cref{fig:auction}.
The samples \ctselect, \ctsort, \haclpolicies, \kremlib, \libsodiumutils,
\opensslutil, \sslcbcrem, \tlslucky, \tlspatched are cryptographic library
code, drawn from
benchmarks for the Binsec/Rel tool~\cite{daniel2020binsec}.
Samples \ctselect, \ctsort and \tlslucky contain known vulnerabilities.
Most are libraries of basic helper routines, except for \sslcbcrem,
\tlslucky and \tlspatched. The latter two are the vulnerable and patched
versions of the infamous ``Lucky13'' TLS vulnerability~\cite{al2013lucky}.
The remaining samples are drawn from the Constant-Time Toolkit (CTTK)
(\url{https://github.com/pornin/CTTK}): \cttkinner is a library of basic helper functions,
\hex is purportedly constant-time routines for converting to/from
binary and hexadecimal strings; \intthirtyone is drawn from big integer library;
\oram is a basic oblivious RAM
(ORAM) library. 

\subsubsection{Accuracy and Bug Discovery.}
For the known vulnerable samples, \Tool and \InferTool correctly detect
the known vulnerabilities. \Tool additionally identifies an out-of-bounds
array access in the big integer library \intthirtyone. This vulnerability
we confirmed by fuzzing the affected code with libFuzzer 
and AddressSanitizer 
 enabled, and was subsequently confirmed
by the developer of the CTTK library. 
\Tool also identified an undocumented information leak
in the \hex CTTK sample, which leaks the location of non-hex characters
in strings. Upon reporting this issue to the developer, we were informed
it was intended behaviour.
This behaviour was also detected by \InferTool.
\Tool identified two information leaks also in the \intthirtyone library
in routines for copying one big integer to another. In particular, if the
destination big integer is not initialised, then these routines can leak
information about the destination memory contents. Limitations in
\InferTool's current implementation prevent it from running on
\intthirtyone at the time of writing.

The information leak identified by \Tool
in \libsodiumutils is similar to that
in \hex and occurs in a routine for converting hex strings to binary,
leaking information if the hex string contains non-hex characters.
Both tools correctly identify the ``Lucky13'' vulnerability in \tlslucky.
\Tool additionally identifies an out-of-bounds array access in this
legacy (now patched) code, heretofore undiagnosed.
The two information leaks that \Tool identifies in the patched ``Lucky13''
code \tlspatched are due to if-conditions that branch on secrets but,
which many compilers optimise away and hence why this sample is
considered to have no known vulnerabilities. Thus whether one regards
these reports as true or false positives depends on
how the code is compiled.

In two samples, \InferTool reports additional information leaks
not reported by \Tool (bold entries).
These arise because \InferTool treats expressions
like \texttt{(a > b) - 1} as if they branch on the boolean condition
\texttt{a > b}. Indeed, \texttt{gcc} 13.1 will compile such
code to a conditional jump when compiled at the lowest optimisation
level~\texttt{-O0} for x86-64,
so we regard these reports as true positives; however we note that on all
higher optimisation levels all modern C compilers will compile such
expressions to straight line code that doesn't leak. 

\subsubsection{Performance.}
\InferTool is orders of magnitude faster than \Tool, in general.
In particular, while \Tool can take minutes to run on some samples,
\InferTool takes no more than a second to analyse each sample.
This
should be expected, for a number of reasons. Firstly, recall that \Tool
uses a timeout mechanism to prune paths during symbolic execution in which
paths are pruned when symbolic execution of individual statements becomes
too slow. On the other hand \InferTool uses a deterministic strategy to
prune paths, by choosing to unroll loops up to a fixed bound only
(by default, once). Thus programs with unbounded loops, like \auction,
take a long time for \Tool to analyse because it keeps unrolling the
main loop until symbolic execution becomes sufficiently slow due to the
growing size of the path condition. This also means that \Tool may
explore loops many more times (and so uncover more behaviours) than
\InferTool in general, so the amount of symbolic execution that the
former performs on a given program is often much greater than the second.
To scale \Tool to the \intthirtyone sample required
increasing its default path pruning timeout. Thus we might expect
that scaling \Tool beyond samples of this size may be challenging.
\InferTool on the other hand suffers no such scalability challenges.

Secondly, \Tool makes use of an external SMT solver in which all
non-spatial assertions are given a relational (i.e.\ two-execution)
encoding to SMT, with very little simplification before formulae are
encoded to SMT.
On the other hand, \InferTool is designed to avoid
the need for relational assertion encoding and in any case uses a
highly performant
in-built satisfiability checking library while continually
performing aggressive
formula simplification. \InferTool benefits from many years of
development effort and optimisation, while having a much simpler
problem to solve (unary symbolic execution). \Tool on the other hand
has far fewer optimisations and has not been designed for speed,
while solving a much harder problem (relational symbolic
execution). 

We note that the analysis times of \InferTool also dwarf the reported
analysis times of the relational symbolic executor
Binsec/Rel~\cite{daniel2020binsec} which, like \Tool, takes minutes to analyse
some samples (e.g.\ the ``Lucky13'' sample for which it requires over
an hour of execution time~\cite[Table III]{daniel2020binsec}).

\section{Related Work and Conclusion}

Our logic \InsecSL is the relational analogue of ISL~\cite{raadlocal},
in the same way that Security Concurrent Separation Logic (\SecCSL)~\cite{Ernst_Murray_19} is the relational analogue of traditional separation logic~\cite{Reynolds_02,OHearn_04}.
\InsecSL can also be seen as the under-approximate dual of \SecCSL,
in the same way that Incorrectness Logic~\cite{OHearn_19} is the under-approximate dual of Hoare logic.
Despite \InsecSL being relational, our symbolic execution procedure is purposefully essentially identical to that for
ISL~\cite{raadlocal,le2022finding}. This allowed us to implement it
as an extension of the existing symbolic execution implementation for
ISL in the Infer tool.

Our symbolic execution procedure is also somewhat similar to
relational symbolic execution~\cite{farina2019relational} (RSE). However,
RSE is not defined for programs with nondeterminism (including from
dynamic memory allocation or external input, both of which we support).
Indeed, RSE was proved sound with respect to over-approximate
Relational Hoare logic~\cite{Benton_04}, whereas ours is based on our
under-approximate logic \InsecSL. We conjecture that extending RSE to
handle nondeterminism would be non-trivial, not least because
over-approximate logics cannot precisely describe errors in nondeterministic
programs (as we noted in \cref{sec:nondet}). Unlike RSE, which is a whole-program analysis, our method is
compositional, allowing it also be applied incrementally.

The recently developed Outcome Logic~\cite{zilberstein2023outcome}
unifies underapproximative and overapproximative reasoning within a uniform framework.
It would be interesting to instantiate this approach with our relational setting.

Declassification is the act of intentionally revealing
sensitive information in a controlled way.
This aspect is orthogonal to the contribution of \InsecSL and could be incorporated
with standard approaches~\cite{banerjee2008expressive}.

We have presented \InsecSL, a logic that soundly discovers insecurities in
program code. The logic strikes a particular balance:
Despite being based on a relational semantic foundation,
it is fairly straight-forward to automate
and inherits many strengths of comparable approaches like ISL,
foremost being compositional.
We have demonstrated that it is capable of precise reasoning
about real insecurities (and errors) in C source code.

\bibliographystyle{splncs04}
\bibliography{references}

\begin{thebibliography}{10}
\providecommand{\url}[1]{\texttt{#1}}
\providecommand{\urlprefix}{URL }
\providecommand{\doi}[1]{https://doi.org/#1}

\bibitem{al2013lucky}
Al~Fardan, N.J., Paterson, K.G.: Lucky thirteen: Breaking the tls and dtls
  record protocols. In: IEEE Symposium on Security and Privacy. pp. 526--540.
  IEEE (2013)

\bibitem{banerjee2008expressive}
Banerjee, A., Naumann, D.A., Rosenberg, S.: Expressive declassification
  policies and modular static enforcement. In: IEEE Symposium on Security and
  Privacy. pp. 339--353. IEEE (2008)

\bibitem{barthe2019formal}
Barthe, G., Blazy, S., Gr{\'e}goire, B., Hutin, R., Laporte, V., Pichardie, D.,
  Trieu, A.: Formal verification of a constant-time preserving c compiler.
  PACMPL  \textbf{4}(POPL),  1--30 (2020)

\bibitem{Benton_04}
Benton, N.: Simple relational correctness proofs for static analyses and
  program transformations. In: POPL. pp. 14--25 (2004)

\bibitem{calcagno2009compositional}
Calcagno, C., Distefano, D., O'Hearn, P., Yang, H.: Compositional shape
  analysis by means of bi-abduction. In: POPL. pp. 289--300 (2009)

\bibitem{Clarkson_Schneider_10}
Clarkson, M.R., Schneider, F.B.: Hyperproperties. Journal of Computer Security
  \textbf{18}(6),  1157--1210 (2010)

\bibitem{daniel2020binsec}
Daniel, L.A., Bardin, S., Rezk, T.: {BINSEC/REL}: Efficient relational symbolic
  execution for constant-time at binary-level. In: IEEE Symposium on Security
  and Privacy. pp. 1021--1038. IEEE (2020)

\bibitem{deVries_Koutavas_11}
De~Vries, E., Koutavas, V.: Reverse hoare logic. In: SEFM. pp. 155--171 (2011)

\bibitem{Eilers2018}
Eilers, M., M{\"u}ller, P., Hitz, S.: Modular product programs. In: ESOP. pp.
  502--529 (2018)

\bibitem{Ernst_Murray_19}
Ernst, G., Murray, T.: \textsc{SecCSL}: Security concurrent separation logic.
  In: CAV. pp. 208--230 (2019)

\bibitem{farina2019relational}
Farina, G.P., Chong, S., Gaboardi, M.: Relational symbolic execution. In: PPDP.
  pp. 1--14 (2019)

\bibitem{le2022finding}
Le, Q.L., Raad, A., Villard, J., Berdine, J., Dreyer, D., O'Hearn, P.W.:
  Finding real bugs in big programs with incorrectness logic. PACMPL
  \textbf{6}(OOPSLA1),  1--27 (2022)

\bibitem{molnar2005program}
Molnar, D., Piotrowski, M., Schultz, D., Wagner, D.: The program counter
  security model: Automatic detection and removal of control-flow side channel
  attacks. In: International Conference on Information Security and Cryptology.
  pp. 156--168. Springer (2005)

\bibitem{Extended}
Murray, T., Yan, P., Ernst, G.: Compositional vulnerability detection with
  insecurity separation logic(extended version) (2023), available online:
  \url{https://covern.org/insecurity.html}

\bibitem{OHearn_04}
O'Hearn, P.W.: Resources, concurrency and local reasoning. In: CONCUR. pp.
  49--67. Springer (2004)

\bibitem{OHearn_19}
O'Hearn, P.W.: Incorrectness logic. PACMPL  \textbf{4}(POPL),  1--32 (2019)

\bibitem{raadlocal}
Raad, A., Berdine, J., Dang, H.H., Dreyer, D., O'Hearn, P., Villard, J.: Local
  reasoning about the presence of bugs: Incorrectness separation logic. In: CAV
  (2020)

\bibitem{Reynolds_02}
Reynolds, J.C.: Separation logic: A logic for shared mutable data structures.
  In: LICS. pp. 55--74. IEEE (2002)

\bibitem{Sabelfeld_Myers_03}
Sabelfeld, A., Myers, A.C.: Language-based information-flow security. IEEE
  Journal on Selected Areas in Communications  \textbf{21}(1),  5--19 (2003)

\bibitem{Yang07}
Yang, H.: Relational separation logic. Theoretical Computer Science
  \textbf{375}(1-3),  308--334 (2007)

\bibitem{zilberstein2023outcome}
Zilberstein, N., Dreyer, D., Silva, A.: Outcome logic: A unifying foundation
  for correctness and incorrectness reasoning. PACMPL  \textbf{7}(OOPSLA1),
  522--550 (2023)

\end{thebibliography}

\ifExtended
\appendix

\section{Appendix}

%

\subsection{Language Semantics}\label{sec:sems}

The small-step semantics for the language over which \InsecSL is defined
is defined in \cref{fig:sems}.

\begin{figure}
\scalebox{0.82}{
  \begin{minipage}{1.2\textwidth}
\begin{mathpar}
  \infer{}{\sems{\Run{\Skip}{s}{h}}{\singlist{\tau}}{\Stop{s}{h}}}

  \infer{}{\sems{\Run{\Input{x}{e}}{s}{h}}{\singlist{\In{\eval{e}{s}}{v}}}{\Stop{\funupd{s}{x}{v}}{h}}}
  
  \infer{}{\sems{\Run{\Assign{x}{e}}{s}{h}}{\singlist{\tau}}{\Stop{\funupd{s}{x}{\eval{e}{s}}}{h}}}

  \infer{a = \eval{p}{s} \and a \not\in \dom{h} \lor h(a) = \bot}{\sems{\Run{\Load{x}{p}}{s}{h}}{\singlist{\tau}}{\Abort{s}{h}}}

  \infer{a = \eval{p}{s} \and h(a) = v}{\sems{\Run{\Load{x}{p}}{s}{h}}{\singlist{\tau}}{\Stop{\funupd{s}{x}{v}}{h}}}

  \infer{a = \eval{p}{s} \and a \not\in \dom{h} \lor h(a) = \bot}{\sems{\Run{\Store{p}{e}}{s}{h}}{\singlist{\tau}}{\Abort{s}{h}}}

  \infer{a = \eval{p}{s} \and h(a) = v}{\sems{\Run{\Store{p}{e}}{s}{h}}{\singlist{\tau}}{\Stop{s}{\funupd{h}{a}{\eval{e}{s}}}}}

  \infer{a \not\in \dom{h} \lor h(a) = \bot}{\sems{\Run{\Alloc{x}{e}}{s}{h}}{\singlist{\Allocate{a}}}{\Stop{\funupd{s}{x}{a}}{\funupd{h}{a}{\eval{e}{s}}}}}

  \infer{a = \eval{p}{s} \and h(a) = v}{\sems{\Run{\Free{p}}{s}{h}}{\singlist{\tau}}{\Stop{s}{\funupd{h}{a}{\bot}}}}

  \infer{a = \eval{p}{s} \and a \not\in \dom{h} \lor h(a) = \bot}{\sems{\Run{\Free{p}}{s}{h}}{\singlist{\tau}}{\Abort{s}{h}}}

  \infer{}{\sems{\Run{\Output{e_\ell}{e}}{s}{h}}{\Out{\eval{e_\ell}{s}}{\eval{e}{s}}}{\Stop}{s}{h}}
  
  \infer{\sems{\Run{c_1}{s}{h}}{\sigma}{\Abort{s'}{h'}}}
        {\sems{\Run{\Seq{c_1}{c_2}}{s}{h}}{\sigma}{\Abort{s'}{h'}}}

  \infer{\sems{\Run{c_1}{s}{h}}{\sigma}{\Stop{s'}{h'}}}
        {\sems{\Run{\Seq{c_1}{c_2}}{s}{h}}{\sigma}{\Run{c_2}{s'}{h'}}}

  \infer{\sems{\Run{c_1}{s}{h}}{\sigma}{\Run{c_1'}{s'}{h'}}}
        {\sems{\Run{\Seq{c_1}{c_2}}{s}{h}}{\sigma}{\Run{\Seq{c_1'}{c_2}}{s'}{h'}}}

        \\
   \infer{\eval{b}{s} = \true}{\sems{\Run{\ITE{b}{c_1}{c_2}}{s}{h}}{\singlist{\tau}}{\Run{c_1}{s}{h}}}

   \infer{\eval{b}{s} \not= \true}{\sems{\Run{\ITE{b}{c_1}{c_2}}{s}{h}}{\singlist{\tau}}{\Run{c_2}{s}{h}}}

   \infer{\eval{b}{s} = \true}{\sems{\Run{\While{b}{c}}{s}{h}}{\singlist{\tau}}{\Run{\Seq{c}{\While{b}{c}}}{s}{h}}}

   \infer{\eval{b}{s} \not= \true}{\sems{\Run{\While{b}{c}}{s}{h}}{\singlist{\tau}}{\Run{\Skip}{s}{h}}}
   
   \infer{\eval{b}{s} = \true}{\sems{\Run{\Assume{b}}{s}{h}}{\singlist{\tau}}{\Stop{s}{h}}}
\end{mathpar}
\end{minipage}
}
\caption{Small step semantics of the language for \InsecSL.\label{fig:sems}
  For a function~$f$ we write~$\funupd{f}{x}{v}$ to denote function update, i.e.\
  to abbreviate the function that
  behaves like~$f$ everywhere except for the argument~$x$ for which it
  returns~$v$.}
\end{figure}

  \subsection{Assertion Semantics}\label{sec:insecsl-assertion-sem}

  \begin{figure}[h]
    Using the abbreviations:
  \[
  \begin{array}{r@{\,}c@{\,}l}
   (s,h) \vDash \pto{e_p}{e_v} &\iff& h = \{\eval{e_p}{s} \mapsto \eval{e_v}{s}\} \\
    (s,h) \vDash \pinvalid{e_p} &\iff& h = \{\eval{e_p}{s} \mapsto \bot\}
  \end{array}
  \] 
  \[
  \begin{array}{r@{\,}c@{\,}l}
  (s,h)\ (s',h') \vDash_{\ell} e           &\iff& \eval{e}{s} = \kw{true} \land \eval{e}{s'} = \kw{true} \land h = h' = \emptyset\smallskip \\
  (s,h)\ (s',h') \vDash_{\ell} \Sec{e}{e_\ell} &\iff& \eval{e_\ell}{s} = \eval{e_\ell}{s'} \land (\eval{e_\ell}{s} \sqsubseteq \ell \implies \eval{e}{s} = \eval{e}{s'}) \land h = h' = \emptyset\smallskip \\
  (s,h)\ (s',h') \vDash_{\ell} \Insec{e}{e_\ell} &\iff& \eval{e_\ell}{s} = \eval{e_\ell}{s'} \land \eval{e_\ell}{s} \sqsubseteq \ell \land \eval{e}{s} \not= \eval{e}{s'} \land h = h' = \emptyset \smallskip \\
  (s,h)\ (s',h') \vDash_{\ell} \Emp &\iff& h = h' = \emptyset \smallskip \\
  (s,h)\ (s',h') \vDash_{\ell} \pto{e_p}{e_v} &\iff& (s,h) \vDash \pto{e_p}{e_v} \land (s',h') \vDash \pto{e_p}{e_v} \smallskip \\
  (s,h)\ (s',h') \vDash_{\ell} \pinvalid{e_p} &\iff& (s,h) \vDash \pinvalid{e_p} \land\  (s',h') \vDash \pinvalid{e_p} \smallskip \\
    (s,h)\ (s',h') \vDash_{\ell} P_1 \star P_2 &\iff& \ \text{there are disjoint subheaps}\ h_1, h_2, \ \text{and}\  h_1', h_2' \\
    && \text{where}\ h = h_1 \uplus h_2 \land h' = h_1' \uplus h_2' \\
    && \text{such that}\ (s,h_1)\ (s',h_1') \vDash_{\ell} P_1\ \text{and}\ (s,h_2)\ (s',h_2') \vDash_{\ell} P_2 \smallskip \\
    (s,h)\ (s',h') \vDash_{\ell} \exists x.\ P\ x &\iff& \text{there are values}\ v,\ v'\ \\ && \text{such that}\ (s(x:=v),h)\ (s'(x:=v'),h') \vDash_{\ell} P \smallskip \\
  (s,h)\ (s',h') \vDash_{\ell} P \implies Q &\iff& (s,h)\ (s',h') \vDash_{\ell} P \ \text{implies}\ (s,h)\ (s',h') \vDash_{\ell} Q \smallskip \\
  (s,h)\ (s',h') \vDash_{\ell} \kw{false} && \text{never}
  \end{array}
  \]
  \caption{Semantics of \InsecSL assertions.\label{fig:rel-sems}}
  \end{figure}
  
  The semantics of \InsecSL assertions are given in \cref{fig:rel-sems}.
  Most of these are familiar and inherited from their counterparts in
  \SecCSL~\cite{Ernst_Murray_19}. As in \SecCSL, \InsecSL assertions
  are given a relational semantics~\cite{Yang07}, i.e.\ are evaluated against a pair
  of states~$(s,h),\ (s',h')$. We write $(s,h)\ (s',h') \vDash_{\ell} P$ to
  mean that assertion~$P$ holds in the pair of states $(s,h)\ (s',h')$.
  The security level~$\ell$ denotes the
  security level of the attacker (see \cref{sec:attacker}).

  Implication and \kw{false} are lifted in the obvious way. $\exists\ x.\ P\ x$
  holds when a pair of values~$v$, $v'$ can be found for~$x$ in
  the left and right states respectively to make~$P$ hold.
  Pure expressions~$e$ are given a boolean interpretation by testing whether
  they evaluate to a distinguished value~$\kw{true}$ in both states.
  Similarly, spatial assertions like~$\Emp$, $\pto{e_p}{e_v}$ and~$\pinvalid{e_p}$
  essentially assert the standard separation logic assertion semantics
  over both states. Separating conjunction lifts its ordinary separation
  logic counterpart over pairs of states:
  $P_1 \star P_2$ holds when each heap can be partitioned into a left and
  right part,
  so that $P_1$ holds of the two left parts, and $P_2$ does likewise for
  the two right parts.

  The semantics of $\Sec{e}{e_\ell}$ remain unchanged from \SecCSL, and
  assert that~$e$ is known to the attacker if the attacker is able to
  observe~$e_\ell$-level outputs or, equivalently, $e$ is known to
  the attacker if the attacker's level is greater than or equal to that denoted by~$e_{\ell}$. Recall that $\ell$ denotes the attacker's security level.
  We say that in a pair of
  states the attacker knows the value of some expression~$e$, if $e$ evaluates
  to identical values in those two states. Thus
  $\Sec{e}{e_\ell}$ holds between two states
  precisely when, if the level denoted by~$e_\ell$ is observable to
  the attacker ($\eval{e_\ell}{s} \sqsubseteq \ell$),
  the two states agree on the value of~$e$.

  Agreement on~$e$ between the two states formalises that the attacker
  knows~$e$. For this reason, \emph{disagreement} on~$e$ formalises that
  the attacker has some uncertainty about~$e$. Hence, the semantics
  for~$\Insec{e}{e_\ell}$.

  \subsection{Extending \InsecSL to Constant-Time Security}\label{sec:ct}

We noted earlier in \cref{sec:attacker} that the security property and
attacker model targeted by \InsecSL is weaker than that of
\emph{constant-time security}~\cite{barthe2019formal}. \InsecSL forbids
a program to explicitly output or  branch on secrets. Constant-time security
additionally forbids a program from performing secret-dependent memory
accesses.

Extending \InsecSL to
constant-time security is straightforward. We briefly
sketch how. Doing so
adds additional rules for loading and storing to the heap to detect
insecurity. Similarly to \textsc{OutInsec}, these rules have in
their result that the pointer~$p$ being loaded from (respectively stored to)
is not known to the attacker: $\Insec{p}{\ell}$. The existing OK rules
have the converse added to their results: $\Sec{p}{\ell}$.

The semantics of the language (\cref{sec:sems})
is extended to record in the schedule~$\sigma$ the address of each pointer
that is loaded from and stored to, effectively making these outputs of
the program. The security
property then imposes the extra requirement that in the two executions,
these addresses are
identical.

Soundness then follows from a similar argument as that for the existing output
rules.

\fi 
\end{document}